 \documentclass[apj,numberedappendix]{emulateapj}

\def\simlt{\lower.5ex\hbox{$\; \buildrel < \over \sim \;$}}
\def\simgt{\lower.5ex\hbox{$\; \buildrel > \over \sim \;$}}

\def\beq{\begin{equation}}
\def\eeq{\end{equation}}
\def\ba{\begin{eqnarray}}
\def\ea{\end{eqnarray}}
\def\bB{{\,\mathbf B}}

\def\Sect{{\rm Section}}

\def\Eq{Equation}

\def\tauT{\tau_{\rm T}}
\def\Urad{U_{\rm rad}}
\def\lrad{\ell_{\rm rad}}
\def\lB{\ell_B}
\def\sT{\sigma_{\rm T}}

\def\TC{T_{\rm C}}

\def\EIC{E_{\rm IC}}

\def\ls{\ell_s}
\def\wmax{w_{\max}}
\def\tC{t_{\rm IC}}
\def\ts{t_{\rm syn}}
\def\rL{r_{\rm L}}
\def\eps{\epsilon}
\def\taugg{\tau_{\gamma\gamma}}
\def\sgg{\sigma_{\gamma\gamma}}

\def\vrec{v_{\rm rec}}
\def\brec{\beta_{\rm rec}}

\def\tdec{t_{\rm drag}}

\def\gav{\bar{\gamma}_e}
\def\Esyn{E_{\rm syn}}

\def\gabs{\gamma_{\rm abs}}
\def\fdrag{f_{\rm drag}}
\def\fpush{f_{\rm push}}
\def\Ukin{U_{\rm kin}}
\def\taupl{\tau_{\rm pl}}
\def\ypl{y_{\rm pl}}

\def\tann{t_{\rm ann}}
\def\tres{t_{\rm res}}
\def\lcr{\ell_{\star}}
\def\fHE{f_{\rm HE}}
\def\taucr{\tau_{\star}}
\def\fKN{\zeta_{\rm KN}}
\def\amin{a_{\min}}
\def\amax{a_{\max}}
\def\Lsyn{L_{\rm syn}}
\def\LIC{L_{\rm IC}}
\def\LHE{L_{\rm HE}}

\def\Ueff{U_{\rm eff}}
\def\muabs{\mu_{\rm abs}}

\def\Lsc{L_{\rm sc}}
\def\Lann{L_{\rm ann}}
\def\tpush{t_{\rm push}}

\newbox\grsign \setbox\grsign=\hbox{$>$} \newdimen\grdimen \grdimen=\ht\grsign
\newbox\simlessbox \newbox\simgreatbox \newbox\simpropbox
\setbox\simgreatbox=\hbox{\raise.5ex\hbox{$>$}\llap
     {\lower.5ex\hbox{$\sim$}}}\ht1=\grdimen\dp1=0pt
\setbox\simlessbox=\hbox{\raise.5ex\hbox{$<$}\llap
     {\lower.5ex\hbox{$\sim$}}}\ht2=\grdimen\dp2=0pt
\setbox\simpropbox=\hbox{\raise.5ex\hbox{$\propto$}\llap
     {\lower.5ex\hbox{$\sim$}}}\ht2=\grdimen\dp2=0pt
\def\simgt{\mathrel{\copy\simgreatbox}}
\def\simlt{\mathrel{\copy\simlessbox}}

\begin{document}

\title{Radiative magnetic reconnection near accreting black holes} 

\author{Andrei M. Beloborodov}
\affil{Physics Department and Columbia Astrophysics Laboratory,
Columbia University, 538  West 120th Street New York, NY 10027;
amb@phys.columbia.edu}

\begin{abstract}
A radiative mechanism is proposed for magnetic flares near luminous accreting 
black holes. It is
based on recent first-principle simulations of magnetic reconnection, which show 
a hierarchical chain of fast-moving plasmoids. The reconnection occurs in a compact region 
(comparable to the black hole radius), and the chain experiences fast Compton cooling 
accompanied by electron-positron pair creation.
The distribution of plasmoid speeds is shaped by radiative losses,
and the self-regulated chain radiates its energy in hard X-rays. 
The mechanism is illustrated by Monte-Carlo simulations of the transfer of seed soft 
 photons through the reconnection layer.
The emerging radiation spectrum has a cutoff near 100~keV similar to
the hard-state spectra of X-ray binaries and AGN.
We discuss how the chain cooling differs from previous phenomenological 
emission models, and suggest that 
it can explain the hard X-ray activity of accreting black holes from first principles. 
Particles accelerated at the X-points of the chain produce an additional 
high-energy component, explaining ``hybrid Comptonization'' observed in Cyg~X-1. 
\end{abstract}

\keywords{
accretion, accretion disks ---
magnetic reconnection ---
radiation mechanisms: general ---
relativistic processes ---
stars: black holes ---
galaxies: active 
}


\section{Introduction}

Accretion disks around black holes are bright sources of X-rays.
They are observed to radiate in ``soft'' and ``hard'' states 
(e.g. \citealt{Zdziarski_Gierlinski2004}). The soft state is 
dominated by quasi-thermal emission from an optically thick accretion disk, and the 
hard state is dominated by hard X-rays that come from a plasma of 
a moderate optical depth. A canonical example is Cyg~X-1. 
Its hard-state spectrum is usually explained by a phenomenological model of
Comptonization of soft X-rays in a hot plasma (``corona'') with electron temperature 
$kT_e\sim 100$~keV. The weak MeV tail of the observed spectrum is explained by 
the presence of additional nonthermal electrons \citep{Coppi1999,McConnell+2002}.

This phenomenological model invokes an unspecified heating mechanism that 
balances the fast Compton cooling. The thermal electrons are cooled by seed soft 
photons through unsaturated Comptonization (e.g. \citealt{Rybicki_Lightman1979}), 
which satisfies the condition
\beq
   y\sim 4\,\frac{kT_e}{m_ec^2}\,\tauT^2\sim 1,
\eeq
where $\tauT\simgt 1$ is the (Thomson) scattering optical depth of the heated plasma,
and $m_ec^2=511$~keV is the electron rest-mass energy. This condition, together with 
$kT_e\sim 100$~keV, implies $\tauT\sim 1$. Similar temperatures and optical depths 
are also inferred from observations of AGN \citep{Fabian+2015}. 

Why the heated plasma has the optical depth  $\tauT\sim 1$ is still an open issue.
One possibility is that $\tauT$ is regulated by creation of $e^\pm$ pairs in 
photon-photon collisions \citep{Guilbert+1983,Svensson1987,Stern+1995}.
A remarkable feature of observed spectra is the sharpness of the cutoff near 100~keV; 
in particular, the hard-state luminosity of Cyg~X-1 drops by a factor $\sim 30$ between 
200~keV and 1~MeV. This is often interpreted as evidence for Comptonization by 
nearly isothermal plasma, with an exponentially suppressed electron population at 
energies $E_e\gg kT_e\approx 100$~keV.
It is, however, unclear why the emission region is nearly isothermal, as
$T_e$ must reflect the heating rate, which varies in space and time.

Attempts to develop an emission model from first principles, avoiding phenomenological 
assumptions, must be based on a concrete mechanism of energy release.
A plausible mechanism is magnetic reconnection \citep{Galeev+1979}. 
It occurs in current sheets formed by magnetic loops above 
the accretion disk, resembling solar activity. Unlike solar flares, 
the current sheets are generated mainly by the disk rotation.
Differential rotation of the magnetic loop footpoints (one of which may be on 
the black hole) leads to inflation and opening of the loop, with a current sheet 
separating the two opposite open magnetic fluxes \citep{Romanova+1998,Parfrey+2015}. 
The energy of the inflated loop is then released through reconnection of the two fluxes.
Another difference from solar flares is that the magnetic energy density in the disk 
corona, $B^2/8\pi$, can strongly dominate over the rest-mass density of the plasma,
$\rho c^2$, i.e. reconnection is relativistic.

A number of ab initio microscopic simulations of relativistic magnetic reconnection have 
been performed using particle-in-cell (PIC) method \citep{Sironi_Spitkovsky2014,
Melzani+2014,Guo+2016,Sironi+2016,Werner+2016}.
The simulations demonstrate a chain of distinct plasmoids in the reconnection layer,
carrying a broad nonthermal electron distribution rather than a Maxwellian plasma. 
The results have been applied to nonthermal emission from relativistic outflows, 
including gamma-ray flares in the Crab nebula and blazars 
\citep{Cerutti+2014,Kagan+2016,Petropoulou+2016}. 

The PIC simulations have not explored yet the regime of dominant radiative losses 
that become inevitable if magnetic reconnection occurs in a compact region near 
a luminous accreting black hole. In this paper, we make analytical estimates of 
radiative losses and their effects on the plasmoid chain.
In particular, inverse Compton scattering and $e^\pm$ pair creation play a key role.
Then we use a Monte-Carlo simulation to evaluate the Comptonized
radiation spectrum produced by a compact magnetic flare with a relativistic plasmoid 
chain. The results appear consistent with the observed hard-state 
spectra of accreting black holes, offering a solution to the puzzle of the 100-keV cutoff.


\section{Basic parameters}

\subsection{Magnetization $\sigma$ and compactness $\ell$}

One dimensionless parameter of the reconnection problem is the magnetization,
\beq
\label{eq:sigma}
   \sigma=\frac{B^2}{4\pi \rho c^2}=\frac{2U_B}{\rho c^2},
\eeq
where $B$ is the magnetic field, $U_B=B^2/8\pi$, and $\rho$ is the mass density of 
the plasma. We will neglect the guide field, i.e. assume that $\bB$ reverses direction 
across the current sheet. We also assume that the plasma before reconnection is cool
(without energy dissipation, the coronal plasma is kept at the Compton temperature of 
the local radiation field, $k\TC\ll m_ec^2$).
Reconnection in a magnetically dominated corona, $\sigma>1$, 
is of main interest for this paper. Then fast magnetosonic waves have the Lorentz 
factor $\gamma\approx\sigma^{1/2}$, and the plasma bulk motions in the reconnection 
layer achieve a similar $\gamma\sim\sigma^{1/2}$ \citep{Lyubarsky2005,Sironi+2016}.
The parameter $\sigma$ also controls the energy release per particle in the flare. 

In the presence of radiative losses, a second dimensionless parameter appears in 
the reconnection problem --- the ``compactness parameter'' $\ell$, which determines
how the timescale for electron radiative cooling compares with the light crossing 
time of the current sheet, $s/c$. In particular, the timescales for inverse Compton 
cooling (with Thomson scattering) and synchrotron cooling are given by
\beq
\label{eq:cool}
  \frac{\tC}{s/c}=\frac{3}{4\gamma_e\,\lrad},  \qquad 
  \frac{\ts}{s/c}=\frac{3}{4\gamma_e\,\lB\sin^2\theta},
\eeq
where $\gamma_e\gg 1$ is the electron Lorentz factor, $\theta$ is the electron 
pitch angle relative to the magnetic field, $s$ is the size of the current sheet, and
\beq
  \lrad=\frac{\Urad\sT s}{m_ec^2}, \qquad \lB=\frac{U_B\sT s}{m_ec^2}.
\eeq
Here $m_e$ is the electron mass and $\sT$ is the Thomson cross section.
In the radiative regime, most of the dissipated magnetic energy is immediately converted 
to radiation, which escapes with speed $\sim c$, as long as the optical depth of the
flare region is not much larger than unity. Then $\Urad c\sim U_B\vrec$ and
\beq
\label{eq:lrad}
   \lrad\sim \frac{\vrec}{c}\,\lB, 
\eeq
where $\vrec\sim 0.1c$ is the reconnection speed \citep{Lyubarsky2005}. 

Radiation receives energy from the reconnection layer through Compton scattering.
Let $U_s$ be the density of soft (``seed'') radiation in the reconnection region, and
\beq
  \ls=\frac{U_s\sT s}{m_ec^2}.
\eeq  
The ratio $\Urad/U_s=\lrad/\ls$ controls the hardness of the Comptonized spectrum
emerging from the reconnection layer. 
$U_s$ may be supplied externally, in particular by radiation from a dense, cool 
accretion disk. It may also be created locally by the dissipation process itself,
due to synchrotron emission from electrons accelerated in the magnetic flare.

\subsection{Magnetic flares near black holes}

It is straightforward to estimate the characteristic $\lB$ for a current sheet 
created near a black hole of mass $M$ accreting with rate $\dot{M}$.
The energy density of the coronal magnetic field rooted in the accretion disk, 
$U_B$, is a fraction of the disk pressure $P$. It is related to the viscous stress 
driving accretion, $\alpha P\approx \dot{M}(GMr)^{1/2}/4\pi r^2H$, where 
$\alpha=0.01-0.1$ is the Shakura-Sunyaev viscosity parameter, and
$H$ is the half-thickness of the disk at a radius $r$.  A conservative 
estimate for $U_B$ is given by 
\beq
   U_B\sim \frac{\dot{M}(GMr)^{1/2}}{4\pi r^2H}.
\eeq
The characteristic radius of the most luminous region is small --- comparable 
to $r_g=2GM/c^2$ --- in particular if the black hole is rapidly rotating.

An accreting black hole with luminosity $L$ has the accretion rate 
$\dot{M}=L/\varepsilon c^2$, where $\varepsilon\sim 0.1$ is the radiative efficiency.
A luminous accreting black hole typically has $L$ varying around $0.1 L_{\rm Edd}$ 
where $L_{\rm Edd}=4\pi GMm_pc/\sT$ is the Eddington limit, and the scale-height 
of its accretion disk is $H\simlt r_g$. This gives a rough estimate for $U_B$ and the 
corresponding compactness parameter of a current sheet of size $s\sim r_g$,
\beq
\label{eq:UB}
   U_B\sim \frac{m_pc^2}{\sT r_g}, \qquad \lB\sim \frac{m_p}{m_e}.
\eeq
Stronger fields may be sustained in the model of magnetically arrested disks 
\citep{Tchekhovskoy+2011}. 
A typical expected value for $\lB$ in bright X-ray binaries or AGN is between 
$10^3$ and $10^4$.

\Eq~(\ref{eq:lrad}) gives the corresponding $\lrad\sim 0.1\lB\sim 10^2-10^3$. It is 
greater than the average compactness $\bar{\ell}_{\rm rad}=F\sT r/m_ec^3$, where 
$F$ is the average radiation flux in the vicinity of the black hole. For instance, the 
luminosity of Cyg~X-1 in the hard state, $L\sim 3\times 10^{37}$~erg~s$^{-1}$, 
corresponds to $\bar{\ell}_{\rm rad}\sim 10$ (assuming that the luminosity emerges 
from a few Schwarzschild radii $r_g$). The coronal activity should occur in 
flares, i.e. the emission is localized in space and time, and the flares have 
the compactness parameter $\lrad\gg\bar{\ell}_{\rm rad}$. Millisecond flares 
with luminosities increased by a factor $\sim 20$ above the average emission have
been detected in Cyg~X-1 \citep{Gierlinski_Zdziarski2003}.

The estimate~(\ref{eq:UB}) gives a characteristic magnetic field,
\beq
\label{eq:B}
   B=(8\pi U_B)^{1/2}\sim 10^8\left(\frac{M}{10M_\odot}\right)^{-1/2} {\rm ~G}.
\eeq
The corresponding timescale for electron gyration, $\omega_B^{-1}=m_ec/eB$, 
is many orders of magnitude shorter than the light-crossing time of the region,
\beq
  \frac{c}{\omega_B r_g}\sim 10^{-11} \left(\frac{M}{10M_\odot}\right)^{-1/2}.
\eeq
The length $c/\omega_B$ is the smallest scale in the reconnection problem. 
The corresponding ion scale in an electron-proton plasma, 
$c/\omega_{B}=m_pc^2/eB$, is larger by the factor of $m_p/m_e$, but still much
smaller than $r_g$.

\subsection{Pair creation and optical depth}

The reconnection layer has the characteristic thickness $h\sim \brec s$ 
(where $\brec=\vrec/c$), and its optical depth is defined by 
\beq
  \tauT=n_e\sT h, 
\eeq
where $n_e$ is the number density of electrons or positrons.
Without $e^\pm$ plasma, the electron density $n_e$ equals the proton density $n_p$.
The optical depth of the electron-proton plasma in a flare with compactness $\lB$ 
and magnetization $\sigma$ may be expressed as
\beq
\label{eq:tauep}
   \tauT^{ep}\sim \frac{2\brec}{\sigma} \frac{\lB}{m_p/m_e}\ll 1.
\eeq

Pair creation is inevitable as soon as the flare spectrum extends above 
$m_ec^2=0.511$~MeV, because the MeV photons convert to $e^\pm$ pairs 
through photon-photon ($\gamma$-$\gamma$) collisions \citep{Guilbert+1983}.
The absorption optical depth $\taugg$ seen by a gamma-ray depends on its energy
$E=\eps m_ec^2$. Photons with $\eps\gg 1$ mainly interact with target photons of 
low energies $\eps_t\sim \eps^{-1}$ (near the threshold of $\gamma$-$\gamma$ reaction). 
Photons of energies just above $\eps\sim 1$ interact with each other, with a 
cross section $\sgg\sim 0.1\sT$. It is convenient to define 
\beq
\label{eq:l1}
 \ell_1=\frac{U_1\sT\,s}{m_ec^2}=\frac{U_1}{U_B}\,\lB,
\eeq
where $U_1$ is the energy density of photons with $\eps\sim 1$.
Only a small fraction $\sim 10^{-2}$ of Cyg~X-1 luminosity is observed
in the MeV band, which suggests $\ell_1\sim 10^{-2}\lrad\sim 2-10$.

Most of the pairs are created inside or near the reconnection layer.
The rate of photon-photon collisions is quickly reduced with distance $z$ from the MeV 
source due to reduced angles between the colliding photons \citep{Beloborodov1999b}.
In addition, the MeV emission is beamed along the reconnection layer,  as will be 
discussed below. The characteristic thickness of the pair-production region,
$z/s\sim 0.2$, is comparable to the thickness of the reconnection layer $h/s\sim \vrec/c$. 

The rate of pair creation by the MeV photons is given by 
\beq
   \dot{n}_+\sim\sgg c\,n_1^2, \qquad \sgg=\eta\sT, \qquad n_1\sim\frac{U_1}{m_ec^2},
\eeq
where $\eta\sim 0.1$ \citep{Svensson1987}. 
The residence time of plasma in the reconnection layer is 
\beq
  \tres\sim \frac{h}{\vrec}\sim \frac{s}{c},
\eeq
and the density of positrons accumulated in the plasma before it is ejected from
the flare is
\beq
   n_+\sim \dot{n}_+\tres\sim \eta\sT n_1^2 s. 
\eeq   
This is a valid estimate for $n_+$ if the annihilation of $e^\pm$ pairs is negligible, 
i.e. if the annihilation timescale $\tann>\tres$. In the opposite case, $n_+$ is 
determined by the annihilation balance,
\beq
   \dot{n}_{\rm ann}=\frac{3}{8}\,\sT c\,n_+ n_-\sim \sgg c\,n_1^2.
\eeq
Then, using $n_-\approx n_+$, one finds $n_+\sim (8\eta/3)^{1/2} n_1$. 
Note that $\tann=n_+/\dot{n}_{\rm ann}=8/3\sT c\,n_+$, and the boundary
between the two cases, $\tann=\tres$, occurs when 
$\ell_1\approx\lcr=(8/3\eta)^{1/2}\approx 5$.

The optical depth of $e^\pm$ pairs of density $n_\pm=2n_+$ is given by
\begin{eqnarray}
    \tauT\sim n_\pm\sT h\sim \frac{16}{3}\,\brec \times
                          \left\{\begin{array}{cl}
    (\ell_1/\lcr)^2,  &   \;\ell_1<\lcr\\
    \ell_1/\lcr,\;        &   \;\ell_1>\lcr 
                                  \end{array}
                          \right.
\end{eqnarray}
The characteristic optical depth $\tauT\sim(16/3)\brec$ is close to unity. 
The expected $\ell_1$ is comparable to $\lcr$ and is unlikely to greatly exceed it;
hence $\tauT\gg 1$ is not expected.

An estimate for pair density may also be formulated assuming that a fraction $\fHE$ 
of the magnetic energy supplied to the reconnection layer converts to high-energy
particles. The particles will produce inverse Compton (IC) emission, in particular if their 
synchrotron losses are suppressed by small pitch angles or by synchrotron self-absorption. 
Practically all emission above a few MeV is blocked by the $\gamma$-$\gamma$ reaction 
and reprocessed to lower energies through creation of secondary pairs. As a result,
a fraction $Y\ll 1$ of the injected particle energy converts to the rest-mass of 
the secondary $e^\pm$ pairs \citep{Svensson1987}. If synchrotron losses are negligible then
$Y\sim 0.1$. The reduction of $Y$ due to synchrotron losses depends on the 
injection energy of the accelerated particles, which will be discussed below.

The resulting pair creation rate in the reconnection layer is 
\beq
    \dot{n}_\pm\sim \frac{Y\fHE U_B}{\tres m_ec^2}.
\eeq
The annihilation balance is approached if $Y\fHE\,\lB>16/3$. Defining $u=(3/16)Y\fHE\,\lB$,
one can express the optical depth of the pair-loaded reconnection layer as
\begin{eqnarray}
\label{eq:tauT}
    \tauT\sim \frac{16}{3}\,\brec \times
                          \left\{\begin{array}{cl}
    u,           &   \;u<1\\
    u^{1/2},  &   \;u>1 
                                  \end{array}
                          \right.
\end{eqnarray}


\section{Radiative plasmoid chain}

\begin{figure*}[t]
\vspace*{-8cm}
\begin{center}
\includegraphics[width=0.9\textwidth]{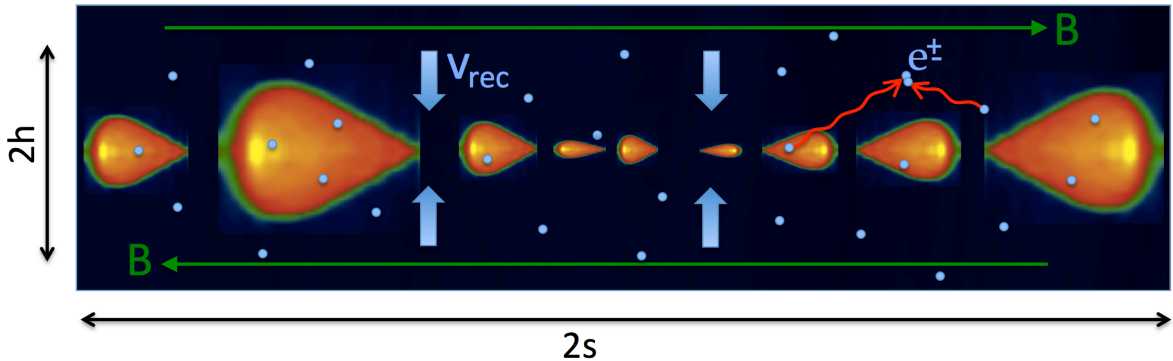} 
\end{center}
\vspace*{-8.1cm}
\caption{Schematic picture of the reconnection layer. Opposite magnetic 
fluxes converge toward the midplane of the layer with velocity $\vrec\sim 0.1c$.
The reconnected magnetic field forms closed islands (plasmoids), which move 
horizontally with various relativistic speeds. Their Lorentz factors $\gamma$ 
reach $\sigma^{1/2}$ where $\sigma$ is the magnetization parameter defined in 
\Eq~(\ref{eq:sigma}). 
The Lorentz factors are controlled by radiative losses and related to 
the plasmoid size $w$ as discussed in \Sect~3.3. 
The plasmoids have a broad distribution of $w$ and $\gamma$, and form 
a self-similar chain. They radiate hard X-rays with a spectrum calculated in \Sect~4.
Photons with energies $E>m_ec^2$ 
convert to $e^\pm$ pairs in photon-photon collisions (shown by the red arrows);
this process greatly increases the optical depth of the reconnection layer.
}
 \end{figure*}

The magnetic flare develops as the current sheet separating two opposite 
magnetic fluxes becomes tearing-unstable, 
which leads to the formation of a chain of plasmoids separated by ``X-points'' 
of the magnetic field lines \citep{Uzdensky+2010}. 
The X-points and plasmoids are continually created with sizes comparable to the 
Larmor radius of the heated particles, $w_{\min}\sim\rL\sim\sigma c/\omega_B$.
The plasmoids merge and grow in size, forming a hierarchical chain
(Figure~1). The maximum plasmoid width, $\wmax$, is a significant fraction of 
the size of the reconnection layer \citep{Loureiro+2012,Sironi+2016}.
``Monster'' plasmoids can grow up to a significant fraction of $r_g$, and so a huge 
range of scales exists in the plasmoid chain, from $\rL$ to $r_g$.

\subsection{Energy deposition}

The PIC simulations show that about half of the magnetic energy advected into 
the reconnection layer is dissipated. The dissipation occurs in two ways: 
\\
(1) A fraction of particles are accelerated by the strong electric fields $E\sim B$ near 
the X-points. The energy gain $E_e$ has the characteristic value of 
$\sim (\sigma/2) mc^2$, comparable to the average energy released per particle
in the magnetic flare ($m=m_e$ for pair plasma and $m=m_p/2$ for hydrogen).  
\\
(2) The plasmoids are accelerated by magnetic stresses to high Lorentz factors 
(up to $\gamma=\sigma^{1/2}$), collide, and merge, dissipating part of their energy.

Recent PIC simulations with large computational boxes show that 
most of the dissipation is associated with hierarchical mergers rather than X-points 
\citep{Sironi+2016}. As reconnection develops and 
$\wmax$ grows, the role of particle acceleration at the X-points is limited to
an initial energy boost of fresh particles, which is followed by energy gains in the mergers.

Simulations of reconnection in a plasma composed of electrons, protons, and $e^\pm$
pairs have not been performed yet. Let 
\beq
  Z=1+\frac{n_\pm}{n_p}
\eeq
be the ``pair loading factor,''
where $n_p$ and $n_\pm$ are the number densities of protons and pairs, respectively.
In an extended range of $1<Z<m_p/m_e$ the plasma rest 
mass is dominated by ions while the particle number is dominated by $e^\pm$ pairs.
Without radiative losses, the behavior of a particle in relativistic magnetic reconnection 
($\sigma\gg 1$) is controlled by its charge and should be independent of its mass.
In particular, the energy gain at an X-point, the corresponding Larmor radius, and the 
subsequent energy gains in mergers must be the same for protons and $e^\pm$,
as indeed observed in electron-ion PIC simulations \citep{Melzani+2014}).

Flares in the magnetically dominated corona, $\sigma\gg 1$, have 
$\tauT^{ep}\ll 1$ (\Eq~\ref{eq:tauep}) and develop $\tauT\sim 1$ due to copious 
pair creation, which implies $Z\gg 1$. It is useful to express $\sigma$ in the form,
\beq
  \sigma=\frac{2U_B}{n_\pm m_ec^2+n_pm_pc^2}
    =\sigma_0\left(1+\frac{n_pm_p}{n_\pm m_e}\right)^{-1},
\eeq 
where 
\beq
   \sigma_0=\frac{2U_B}{n_\pm m_ec^2}=\frac{2\brec\lB}{\tauT}, 
   \qquad \tauT\approx n_\pm\sT h.
\eeq
Assuming that reconnection dissipates about half of the magnetic energy, as
observed in the PIC simulations, the average energy released per particle in 
a plasma with $Z\gg 1$ is $U_B/2n_\pm$. This gives the average electron/positron
Lorentz factor,
\beq
\label{eq:gav}
   \gav\approx\frac{\sigma_0}{4}\approx \frac{\brec\lB}{2\tauT}.
\eeq
Substituting $\brec\sim 0.1$, $\lB\sim m_p/m_e$, and $\tauT\sim 1$, one finds
$\gav\sim 10^2$. This is also the expected characteristic Lorentz factor acquired by particles 
at the X-points.

Energy is also deposited in the chain through plasmoid acceleration by magnetic stresses.
We now estimate how this energy deposition is distributed between plasmoids of 
different sizes $w$. In a self-similar chain, the distribution of plasmoid number $N$ over 
$w$ follows a power law,
\beq
   \frac{dN}{d\ln w}\propto w^{-q}.
\eeq
Here $N$ is counted in a two-dimensional cross section of the reconnection region
(in the reconnection plane $xz$). 
A simplified chain model of \citet{Uzdensky+2010} give $q=1$, and more detailed 
considerations suggest $q\approx 0$ \citep{Huang_Bhattacharjee2012}. Numerical 
simulations by \citet{Sironi+2016} show that $q$ is close to zero on scales $w\ll\wmax$
(the self-similar chain) and increases as $w$ approaches $\wmax$.

The plasmoids are pushed and accelerated by the gradients of magnetic stresses.
Using the plasmoid size $w$ as a characteristic scale to estimate the gradient, one
may write the pushing force (per unit volume) in the form
\beq
\label{eq:fpush}
   \fpush= \xi\,\frac{U_B}{w}.
\eeq
The dimensionless coefficient $\xi<1$ may be found from numerical simulations.
The simulations show that plasmoids accelerate significantly slower than their 
light-crossing time $w/c$, which implies $\xi\ll 1$. \citet{Sironi+2016} find
$\xi\sim \beta\,\brec$ where $\beta=v/c$ is the plasmoid speed. 
This estimate may change when the chain experiences strong Compton drag, 
and this needs to be explored with new PIC simulations. 
Note also that large fluctuations in plasmoid 
motions indicate that $\xi$ has a rather broad distribution around its mean value. 
However, below we only consider the simplest model with a fixed $\xi$.

The work done by $\fpush$ (integrated over the plasmoid cross section in the reconnection 
plane) scales as $w^2 \fpush v$. This can be used to estimate the scaling law for the 
energy deposition into the plasmoid bulk motions in the self-similar chain. 
The deposited power $L$ is distributed over $w$ as
\beq
\label{eq:L}
  \frac{dL}{d\ln w}\propto w^2 \fpush\, v\,\frac{dN}{d\ln w}\propto \xi\,w^{1-q}\,\beta,
\eeq
where $\beta\approx 1$ for relativistic plasmoids. This rough estimate may be refined by 
direct measurements of $dL/d\ln w$ in PIC simulations. For the radiative 
chain model described below (\Sect~3.3) it is important that 
the power deposited into bulk motions is broadly distributed over plasmoid 
size $w$ and, for relativistic plasmoids, $dL/d\ln w$ decreases toward small $w$.

\medskip

\subsection{Cooling of young hot plasmoids}

Electrons and positrons accelerated at the X-points gain Lorentz factors 
comparable to $\bar{\gamma}_e$ on the timescale,
\beq
  t_X\sim\frac{\gav m_ec}{eB} 
      \sim 10^{-9} \left(\frac{M}{10M_\odot}\right)^{-1/2} \frac{r_g}{c}.
\eeq
The shortest possible timescale for cooling is given by
\beq
\label{eq:tsynch1}
   \ts^{\min}\sim \frac{m_e c}{\sT U_B \bar{\gamma}_e}
      \sim 10^{-5}\,\frac{r_g}{c},
\eeq
where we substituted $\gav$ from \Eq(\ref{eq:gav}) and $U_B$ from 
\Eq~(\ref{eq:UB}). One can see that $\ts\gg t_X$, so the accelerated particles 
are pushed into young small plasmoids before they have a chance to radiate their energy.

The cooling time is also longer than the time it takes a nascent plasmoid to develop 
a bulk Lorentz factor $\gamma\gg 1$. Then the Lorentz factor of an electron/positron
in the lab frame may be written as  
\beq
  \gamma_e\sim\gamma\gamma_e^\prime,
\eeq
where $\gamma_e^\prime$ is a ``thermal'' Lorentz factor measured in the plasmoid 
rest frame. Synchrotron losses can reduce only $\gamma_e^\prime$ and occur  
slower if $\gamma\gg 1$. The cooling time in the plasmoid frame 
$\ts^\prime\propto ({B^\prime}^2\gamma_e^\prime)^{-1}$,
and the cooling time in the static lab frame is $\ts=\gamma\ts^\prime$.
Synchrotron self-absorption strongly suppresses cooling when 
$\gamma_e^\prime\simlt 10$.

The timescale for IC cooling in the plasmoid rest frame is given by
\beq
   \tC^\prime\sim \frac{m_ec}{\sT \zeta_{\rm KN}\,\gamma^2 \Urad\gamma_e^\prime},
\eeq
where we used $\Urad^\prime\sim\gamma^2\Urad$, and $\zeta_{\rm KN}<1$ 
describes the reduction due to Klein-Nishina corrections.
An upper bound on $\tC^\prime$ may be estimated as $\tC^\prime\simlt m_ec/\sT\Urad$, 
which gives 
\beq
  \tC^\prime\simlt \frac{10}{\lB}\,\frac{r_g}{c}.
\eeq

The maximum cooling time should be compared with the typical age 
of plasmoids of size $w$: $t_{\rm age}^\prime\sim 10(w/c)$
 (Sironi et al. 2016). This comparison shows that plasmoids 
of sizes $w\gg r_g/\lB$ must be cooled to a non-relativistic temperature. 
The $e^\pm$ pairs trapped in large plasmoids tend to Compton equilibrium with 
radiation, which corresponds to a temperature $k\TC\ll m_ec^2$. Only the ion
component can keep the heat received in reconnection.

This conclusion is not changed by heating due to mergers,
as the mergers of large plasmoids occur slower than their cooling.
In the heating picture discussed by \citet{Sironi+2016},
the energies gained by the plasmoid particles should be proportional to their energies 
before the merger, similar to simple adiabatic heating. Then the hot ions are 
strongly heated while the cooled electrons receive much less energy. 

Since the plasma is dominated by $e^\pm$ pairs ($Z\gg 1$), their cooling
implies losing most of the plasma enthalpy.
As a result, the inertial mass of the plasma is reduced. However,  
the plasmoid does not become much lighter after cooling, because a large fraction 
of its effective mass density is carried by the magnetic field, 
$\rho_{\rm eff}\approx B^2/4\pi c^2$. The strong cooling 
only implies that the plasmoids assume a nearly force-free configuration. 
The force-free hierarchical chains are observed in simulations of reconnection that
neglect the plasma inertia \citep{Parfrey+2013,Parfrey+2015}. They show the same 
X-points, mergers, and the growth of monster plasmoids.
The dynamics of force-free plasmoids are fully controlled by the magnetic stresses in
the chain, which are not much different from those in the PIC models without cooling.

\medskip


\subsection{Compton drag on large plasmoids}

The plasmoid chain has a regular motion component 
toward the exit from the reconnection layer and comparable or faster random motions.
Scattering of ambient photons creates effective viscosity for plasmoids.\footnote{Unlike 
   synchrotron radiation, the IC photons with energies $\EIC^\prime$ in the plasmoid rest 
   frame have a significant average momentum $\sim\EIC^\prime/c$, even when the 
   energetic electrons are isotropic in this frame. The asymmetry of IC 
   radiation is related to the fact that the target soft radiation is beamed in the plasmoid 
   frame due to relativistic aberration. Scattering by electrons moving against the beam 
   is more frequent, leading to preferential IC 
   emission against the beam direction \citep{Odell1981,Phinney1982,Sikora+1996}.
   The bias in the scattering direction causes the plasmoid deceleration.}
As long as scattering occurs in Thomson regime, the drag force per unit volume applied 
to an optically thin plasmoid moving through radiation of density $\Urad$ is given by
\beq
   f_{\rm drag}\approx\beta\gamma^2 {\gamma_e^\prime}^2 \Urad\sT\,n_\pm,
\eeq
where $\gamma=(1-\beta^2)^{-1/2}$ is the bulk Lorentz factor, and
$\gamma_e^\prime$ is the random (``thermal'') Lorentz factor of electrons (or positrons)
in the plasmoid rest frame; $\gamma_e\approx\gamma\gamma_e^\prime$
is the characteristic electron Lorentz factor in the lab frame.
The plasmoid momentum per unit volume is $\sim\beta B^2/4\pi c$,
and the plasmoid deceleration timescale is 
\beq
   \tdec\sim\frac{ \beta B^2}{4\pi c \fdrag}
   \sim\frac{U_B}{\Urad}\,\frac{1}{\gamma^2{\gamma_e^\prime}^2\sT n_\pm c}.
\eeq
Note that $\tdec$ is related to the Compton cooling time $\tC$ by
\beq
  \frac{\tdec}{\tC}\sim \frac{U_B}{\Ukin}, \qquad \Ukin=\gamma_e m_ec^2 n_\pm.
\eeq
In a radiative plasmoid chain, $\Ukin\ll U_B$ and hence $\tdec\gg\tC$, so the plasmoids
are strongly cooled before deceleration. In the strong cooling limit,
$\gamma_e^\prime\approx 1$, $\gamma_e\approx\gamma$, and
the drag force is $\fdrag\approx\beta\gamma^2 \Urad\sT n_\pm$.

The plasmoids are pushed by magnetic forces $\fpush=\xi U_B/w$ (\Sect~3.1)
and capable of accelerating as long as $\fdrag<\fpush$.
The ratio of the two forces may be written as 
\beq
\label{eq:ratio}
  \frac{\fdrag}{\fpush}=\beta\gamma^2\frac{\taupl}{\taucr}, \qquad
  \taucr\equiv\xi\,\frac{U_B}{\Urad}\approx \frac{\xi}{\brec}.
\eeq
Here $\taupl$ is the optical depth of the plasmoid of width $w$,
\beq
   \taupl=n_\pm\sT w.
\eeq
The drag is not important if $\fdrag/\fpush<1$, and from \Eq~(\ref{eq:ratio}) one concludes 
that plasmoids with $\taupl<\taucr$ can become relativistic, $\beta\gamma^2>1$. 
The drag limits their Lorentz factors to $\gamma\approx (\taucr/\taupl)^{1/2}$.
Plasmoids with $\taupl<\taucr/\sigma$ are limited by the ``ceiling'' $\gamma\leq \sigma^{1/2}$, which is not related to drag and observed in the PIC simulations without 
radiative losses \citep{Sironi+2016}. 

Monster plasmoids have sizes comparable 
to the thickness of the reconnection layer $h$, and their optical depths are $\taupl\sim\tauT$.
If $\tauT>\taucr$ (expected for $\xi\ll 1$, see \Sect~3.1) then the drag limits 
the plasmoid speeds even in the nonrelativistic regime $\beta\gamma<1$.
Note also that even without Compton drag, the largest plasmoids in the chain are pushed 
only to mildly relativistic speeds $v\sim 0.5 c$ \citep{Sironi+2016}. 

The bulk motion is saturated, $\fdrag\approx\fpush$, if the duration of the magnetic force, 
$\tpush$, exceeds $\tdec$. In the simplest model, $\tpush$ is the residence time in the 
chain, $\tres\approx s/\beta c$. Using $\Urad/U_B\approx \brec\approx h/s$, the ratio 
$\tdec/\tres$ is conveniently expressed as 
\beq
   \frac{\tdec}{\tres}\sim \frac{\beta}{\gamma^2\tauT}, \qquad \tauT=n_\pm\sT h.
\eeq
One can see that $\tdec<\tres$ for all plasmoids as long as the reconnection layer 
has a substantial optical depth $\tauT\simgt 1$. However, in a realistic chain $\fpush$
varies stochastically on timescales $\tpush\ll\tres$. Therefore, deviations from the force 
balance are expected, and the condition $\fdrag\leq\fpush$ only defines an upper limit for 
the plasmoid speed.

This upper limit still represents a characteristic speed of plasmoids of a given 
optical depth $\taupl$,  
\beq
\label{eq:balance}
   \beta\gamma^2\approx \frac{\taucr}{\taupl}.
\eeq
It can also be expressed using an effective $y$-parameter ---
the product of the scattering probability $\taupl$ (for a soft photon 
propagating through the plasmoid) and the average energy gain in scattering 
$\Delta E/E\sim\beta^2\gamma^2$, 
\beq
  \ypl=a\taupl,  \qquad a\equiv\beta^2\gamma^2.
\eeq
The parameter $\ypl$ is the amplification factor of radiation flowing through the plasmoid.
\Eq~(\ref{eq:balance}), which corresponds to $\fdrag=\fpush$, can be stated as an
energy balance condition: the work done by magnetic forces, $\dot{U}=\fpush v$, 
converts to radiation energy that escapes the plasmoid on the timescale 
$t_{\rm esc}\sim w/c$: $\dot{U}=\ypl U/t_{\rm esc}$.

\Eq~(\ref{eq:L}) described how the power $L$ deposited in the chain is distributed 
over the plasmoid size $w$. For calculations of photon Comptonization 
it is important to know how the power is distributed over the parameter 
$a=\beta^2\gamma^2$, which controls the photon energy gain per scattering.
We will roughly estimate this distribution using \Eq~(\ref{eq:balance}) and considering
chains where the magnetic forces $\fpush$ have the numerical 
coefficient $\xi=const$ (\Eq~\ref{eq:fpush}). Using $\taupl\propto w$,
one finds from \Eq~(\ref{eq:balance}) $a\propto \beta/w$. This gives $w\propto a^{-1}$
for relativistic plasmoids ($a>1$). For non-relativistic plasmoids $a\approx\beta^2$
and one finds $w\propto a^{-1/2}$. In summary, substitution of 
$w\propto \beta/a$ into \Eq~(\ref{eq:L}) gives 
\begin{eqnarray}
\label{eq:La}
    \frac{dL}{d\ln a}\propto 
                          \left\{\begin{array}{cl}
    a^{q/2},      &   \;a<1\\
    a^{q-1},      &   \;a>1 
                                  \end{array}
                          \right.
\end{eqnarray}
The expected range of $0<q<1$ (\Sect~3.1) then implies that most of the power
is deposited into mildly relativistic plasmoids, $a\sim 1$. 
The PIC simulations show $q\approx 0$ for small plasmoids (which have $a\gg 1$);
this implies that the power fraction given to high-$a$ plasmoids decreases as $a^{-1}$.

Monster plasmoids with the maximum $w\sim h$  and $\taupl\sim\tauT$ have the 
minimum $a$. In the drag-limited regime this minimum value is 
\begin{eqnarray}
\label{eq:amin}
    \amin\sim 
                          \left\{\begin{array}{cl}
     (\tauT/\taucr)^{-1/2},    &    \;\tauT<\taucr \\
     (\tauT/\taucr)^{-2},         &   \;\tauT>\taucr
                                  \end{array}
                          \right.
\end{eqnarray}
Note also that the above  estimates assume $\taupl\simlt 1$. 
For optically thick plasmoids, the force balance $\fpush=\fdrag$ is different from 
\Eq~(\ref{eq:balance}) and becomes 
\beq
\label{eq:balance1}
  \beta\gamma^2\approx\taucr, \qquad \taupl>1.
\eeq
The simplest model of saturated plasmoid motion with $\xi=const$ implies
that $\taucr=\xi/\brec\simlt 1$ is a fixed constant; then all optically thick plasmoids 
are moving with the same speed given by \Eq~(\ref{eq:balance1}). 
The actual value of $\xi$ can vary. \citet{Sironi+2016} find $\xi\propto\beta$,
so $\xi$ is reduced for non-relativistic plasmoids.
In addition, $\xi$ (and hence $\taucr$) can be noisy,
and deviations from the force balance, $\fpush=\fdrag$, are
expected. This will affect the statistics of the chain bulk motions, and 
direct PIC simulations should be used to measure in detail the distribution of 
plasmoid speeds in the presence of Compton drag.


\section{Radiated spectrum}

Plasmoids with the broad distribution of Lorentz factors discussed in \Sect~3 upscatter 
soft seed photons and form a Comptonized spectrum. This is the dominant hard X-ray 
component of the magnetic flare, and below we examine it in some detail using a 
Monte-Carlo simulation. Then we discuss additional emission components associated 
with high-energy particles and annihilation of $e^\pm$ pairs.

\subsection{Chain Comptonization}

Let us consider a reconnection layer of a given optical depth $\tauT$. The optical depth 
is dominated by $e^\pm$ pairs; it may vary by a factor of a few around $\tauT\sim 1$ 
depending on the compactness parameter of the flare as discussed in \Sect~2.3.

Scattering of photons will sample the velocity field in the reconnection layer. Let $z=0$ 
be the midplane of the layer; the plasmoids of various sizes move along the $\pm x$ 
directions. The plasma temperature will have nothing to do with Comptonization. 
To emphasize this point, in the simulation we will intentionally take the 
cold-plasma limit $T_e=0$ (more realistically, the cooled plasma relaxes to the 
Compton temperature $k\TC\sim 10$~keV, which weakly affects the results). 

Our simplified Monte-Carlo simulation is set up as follows. We inject soft photons with 
energies $E_s$ in the midplane of the reconnection layer and follow their propagation 
and (multiple) scattering until they reach the boundary $|z|=h$. The layer is approximated 
as an infinite slab filled with plasma of uniform density $n_\pm$. The parameter $\tauT$ is 
defined as $\tauT=n_\pm\sT h$. 

We follow the propagating photon with a small timestep $\Delta t\ll h/c$. 
At each step, the local plasma velocity is assumed to be a random variable.
The photon can be scattered by plasmoids (region I)
or by the plasma between the plasmoids (region II) which converges toward $z=0$ 
with speed $\vrec$. The two regions contain equal amounts of plasma
(by definition of the thickness $h$ of the reconnection layer).\footnote{Scattering at $|z|>h$ 
    will be neglected in the simulation presented below, assuming that the pair density 
    quickly decreases at $|z|>h$ (\Sect~2.3).}
Therefore, our Monte-Carlo simulation will assume that the average column densities 
of regions I and II are equal, i.e. the probabilities for scattering in regions I and II are 
equal as long as scattering occurs with Thomson cross section. The actual (Klein-Nishina)
cross section in the simulation depends on the photon energy and the local plasma velocity. 
In region~II, the velocity is fixed at ${\mathbf v}_{\rm rec}=(0,0,-\vrec {\rm sign}(z))$
with $\vrec=0.1c$. In region~I, the plasmoid speed is drawn from the distribution 
discussed in \Sect~3. Then the scattering probability during timestep $\Delta t$ is 
determined according to the Klein-Nishina cross section.

It is convenient to deal with the distribution of $a=\beta^2\gamma^2$ instead of speed 
$\beta$. In the chain, the value of $a$ is drawn randomly with a probability distribution 
$dP=f(a)\,da$. Thus a single function $f(a)$ encapsulates the chain behavior in our
simulation. It is found from the condition $d\Lsc/d\ln a=dL/d\ln a$, where $\Lsc$ is the 
power gained by photons through scattering and $L$ is the power deposited in the chain, 
e.g. given by \Eq~(\ref{eq:La}). 

Energy extracted from the chain in one scattering of 
a photon of energy $E$ is $\Delta E\approx a E$, and the total extracted power is 
\beq
  \Lsc=\dot{N}_{\rm sc}\int f(a)\,\Delta E\;da,
\eeq
where $\dot{N}_{\rm sc}$ is the total scattering rate in the chain.
This implies $d\Lsc/da\propto a\,f(a)$ and 
\begin{eqnarray}
\label{eq:fa}
  f(a)\propto a^{-2}\frac{dL}{d\ln a} \propto 
                          \left\{\begin{array}{cl}
    a^{\psi_1},      &   \;a<1\\
    a^{\psi_2},         &   \;a>1 
                                  \end{array}
                          \right.
\end{eqnarray}
This relation simply states that the power deposited in a plasmoid with a given $a$
is radiated through scattering with this $a$. 
Accurate values of $\psi_1$ and $\psi_2$ should be taken from PIC 
simulations of radiative reconnection. The estimate in \Eq~(\ref{eq:La})
gives $\psi_1=q/2-2$ and $\psi_2=q-3$. The exact value of $q\sim 0$ is not critical,
and we will fix $q=0$ (we also calculated models with $q=1/2$, with similar results). 
The choice of $\psi_1=-2$ and $\psi_2=-3$ may not be accurate, in particular for 
$\psi_1$. However, the key feature of $f(a)$ --- the break at $a\approx 1$ --- 
should be robust, based on the PIC simulations and the estimates of Compton drag.

The constant of proportionality in \Eq~(\ref{eq:fa}) is determined by the normalization of 
the distribution function $\int f(a)\,da=1$. The distribution extends over the range,
\beq
  \amin<a<\amax.
\eeq
The choice of $\amax\sim\sigma$ weakly affects the results as long as $\amax\gg 1$;
we fix $\amax=100$. The value of $\amin$ will be adjusted to give a desired Compton 
amplification factor $A$. The factor $A=\bar{E}_{\rm esc}/E_s$ is defined as the average 
net energy gain of a photon between its injection with energy $E_s$ and escape with 
energy $E_{\rm esc}$; it is directly calculated in the Monte-Carlo 
simulation.\footnote{The amplification factor $A$ is determined by $\amin$ and $\tauT$.
      This dependence is controlled by the shape of $f(a)$, in particular the choice of 
      $\psi_1$ in \Eq~(\ref{eq:fa}). In our default model, we find numerically
      that $A=10$ corresponds to  $\amin\approx 0.7(\tauT/1.5)^{-1.6}$.}
In our simulations, the photons are injected with energy $E_s$ drawn from a Planckian 
distribution with temperature $kT_s=10^{-3}m_ec^2$.
The three parameters that control the shape of the emerging spectrum are $T_s$, $\tauT$, 
and $A$.

Figure~1 shows the results for $A=10$ and a few values of $\tauT=0.7$,
1.5, and 3. According to \Eq~(\ref{eq:tauT}) these values correspond to $u\approx 1.7$,
8, and 31, and the range $1.7<u<31$ can correspond to $5\times 10^2<\lB<10^4$ if
$Y\fHE\sim 10^{-2}$. One can see in the figure that in all three cases the emerging
spectrum has a hard slope and a sharp cutoff around 100~keV. The cutoff position 
moves from $\sim 200$~keV to $\sim 50$~keV as the optical depth increases from 0.7 
to 3, which corresponds to $\lB$ increasing by a factor of 20 (if $Y\fHE$ is assumed 
to be the same in the three models).

\begin{figure}[t]
\includegraphics[width=0.47\textwidth]{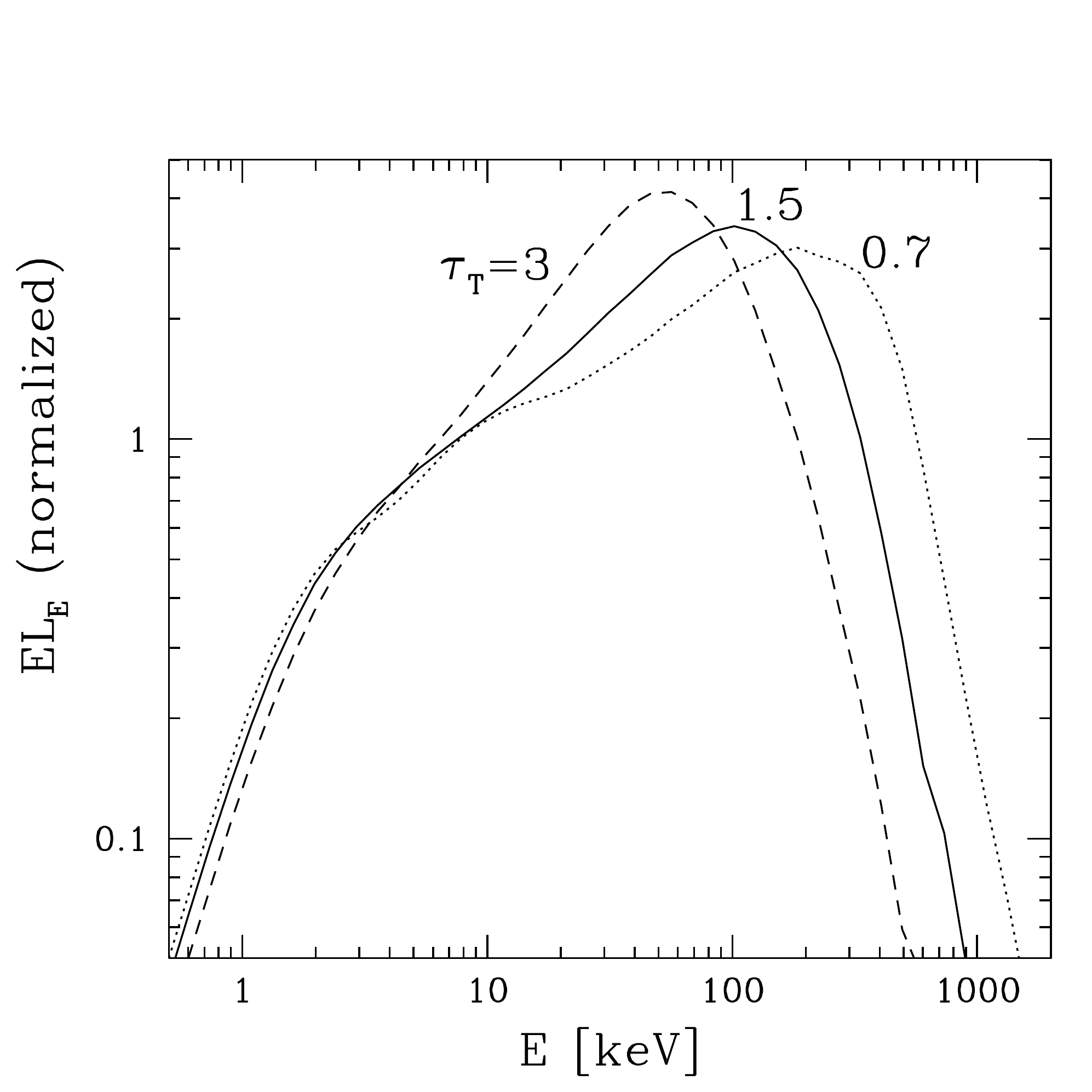} 
\caption{Spectrum emerging from the reconnection layer with Compton 
amplification factor $A=10$. Three models are shown with optical depths
$\tauT=0.7$, 1.5, and 3. The range of $0.7<\tauT<3$ roughly corresponds 
to magnetic compactness varying by a factor of 20 around $\lB\sim m_p/m_e$.
In all the models, the temperature of injected soft radiation is fixed at 
$kT_s=10^{-3}m_ec^2$.
Only the chain bulk Comptonization is simulated in the model. Additional emission 
from particles accelerated at the X-points will create a high-energy 
component that should emerge at $E\simgt m_ec^2$ and carry a few percent of the 
total energy budget (depending on $\fHE$); 
it is not shown in the figure and requires more advanced simulations.}
 \end{figure}

\begin{figure}[t]
\includegraphics[width=0.47\textwidth]{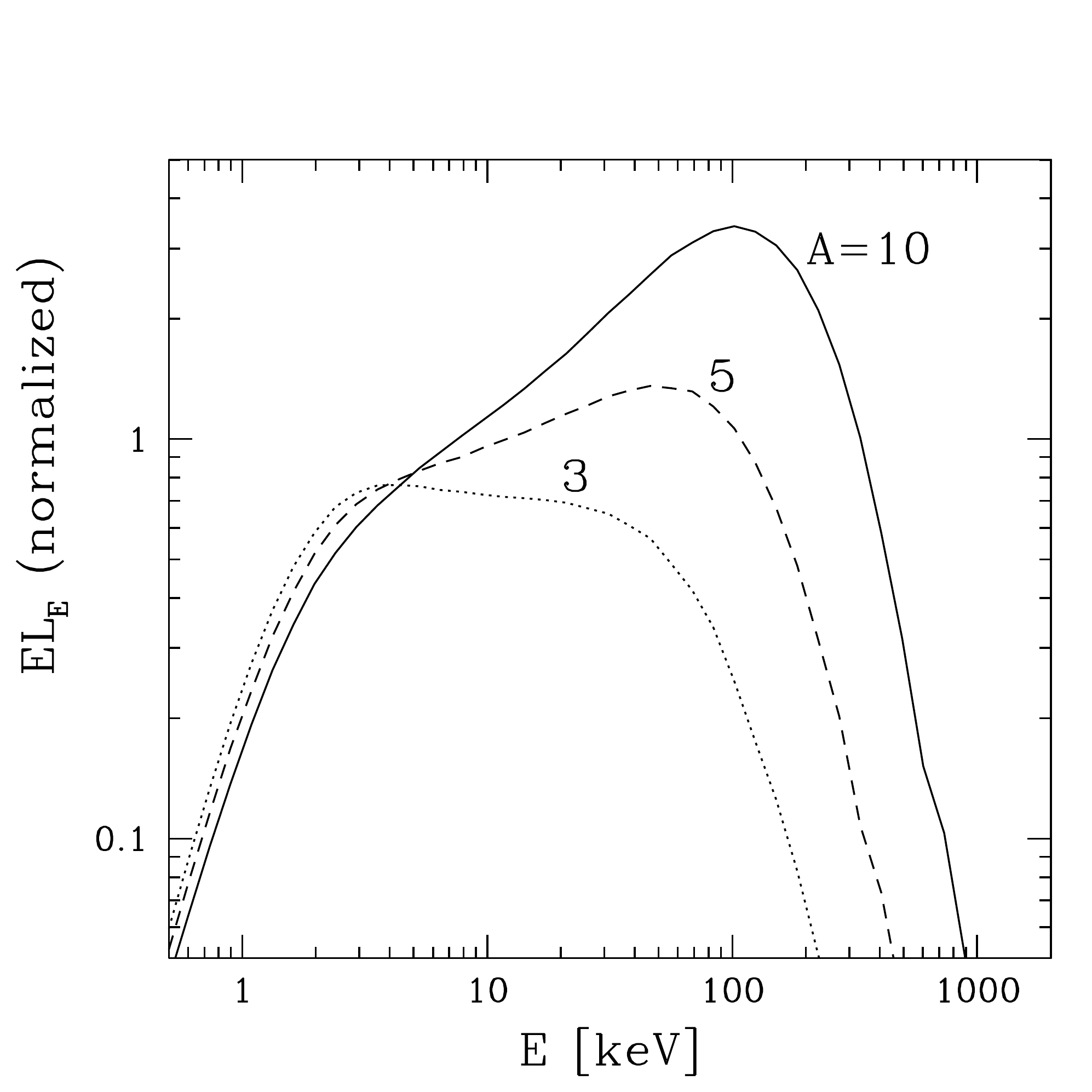} 
\caption{Spectrum emerging from the reconnection layer with Thomson optical depth 
$\tauT=1.5$ in three models with Compton amplification factors $A=3$, 5, and 10.
}
 \end{figure}

The cutoff position --- a fraction of $m_ec^2=511$~keV --- is the result of three effects. 
(1) Most of the chain power is given to mildly relativistic plasmoids
($\gamma\beta\sim 1$) which dominate Comptonization.
(2) Klein-Nishina effects (electron recoil in scattering) 
become strong at photon energies $E>100$~keV,
which reduces emission at these energies. (3) Pair plasma in flares near black holes
never develops a large optical depth $\tauT\gg 1$. The optical depth slowly grows with 
compactness $\lB$, which does not exceed $10^4$. This behavior of $\tauT$
is similar to the known property of thermal $e^\pm$ plasma in annihilation balance 
\citep{Svensson1984,Stern+1995}.  

Photon upscattering by the plasmoid chain  in many respects
resembles thermal Comptonization. In particular, the flare spectrum depends on its 
Compton amplification factor $A$, which is controlled by how many soft photons 
are supplied to the reconnection region.
The Comptonized spectrum becomes significantly softer if the photon supply 
is increased by a factor $\sim 3$, so that the dissipated power per injected photon 
is reduced, giving $A=3$ instead of 10 (Figure~3). This softening is expected, 
as a smaller $A$ means that the probability for a scattered photon to reach high 
energies is reduced. 
Similar to standard thermal Comptonization, the relation between $A$ and the 
spectral slope depends on $T_s$. The slope is softer for AGN, because their  
$T_s$ is lower (see \citealt{Beloborodov1999a} for a quantitive relation).

\subsection{Nonthermal emission from young plasmoids}

In addition to bulk Comptonization by the chain of cooled plasmoids, there 
is radiation from high-energy particles injected in young small plasmoids (\Sect~3.2). 
The characteristic Lorentz factor of these particles before their
cooling is given by \Eq~(\ref{eq:gav}); it is comparable to a few hundred.
They generate synchrotron and high-energy IC emission with power $\LHE$ that 
is a fraction $\fHE$ of the total dissipated power,
\beq
   \LHE=\Lsyn+\LIC=\fHE L.
\eeq
The transformation of the injected particle power $\LHE$ into synchrotron and IC 
emission occurs in a few steps, because 
the IC photons with high energies $\EIC\simlt\gamma_e m_ec^2$ do not escape and 
convert to secondary $e^\pm$ pairs outside the small plasmoids. The power of the 
$e^\pm$ cascade may compete with the synchrotron luminosity of the plasmoid itself,
because its relativistic bulk motion reduces the synchrotron losses (\Sect~3.2).

The IC emission from the plasmoid is beamed along the chain direction (the $x$-axis) 
within the angle $\theta\sim\gamma^{-1}$. Therefore, the secondary pairs are injected 
with angles $\sim\theta$ with respect to the $x$-axis, which is also the direction of the 
background magnetic field. As a result, many secondary pairs are injected with modest 
pitch angles with respect to $\bB$ and their synchrotron cooling is reduced by the 
factor of $\sin^2\theta$.
On the other hand, their IC cooling is reduced by a factor $\fKN<1$ due to the 
Klein-Nishina reduction in the scattering cross section. The ratio of IC and 
synchrotron luminosities generated by the secondary pairs may be estimated as
$\Urad\fKN/U_B\sin^2\theta$.

The high-energy IC photons generate a nonthermal cascade with a pair yield $Y\simlt 0.1$,
contributing to the pair loading of the reconnection layer (\Sect~2.3). In contrast, 
the synchrotron photons are soft, and contribute to the seed radiation for Comptonization 
in the chain. The characteristic energy of synchrotron photons emitted by the secondary 
$e^\pm$ pairs is given by
\beq
   \Esyn\sim 0.3\gamma_e^2\hbar\omega_B
        \sim 10\, B_8 \sin\theta \left(\frac{\gamma_e}{100}\right)^2 {\rm~keV}.
\eeq
Some of the pairs will be created inside large plasmoids, which occupy a significant 
volume in the reconnection layer. Their magnetic fields are far from the 
background pre-reconnection field, and the injected pairs will have large pitch 
angles $\theta\sim 1$. These pairs may give the highest $\Esyn$. In any case, 
$\Esyn$ is unlikely to exceed 10~keV and may be much lower, especially taking into 
account that the secondary pairs likely have $\gamma_e<100$. 

For sufficiently low $\gamma_e<\gabs$, synchrotron self-absorption suppresses 
synchrotron emission. The value of $\gabs$ can be estimated using the absorption 
coefficient $\muabs$ \citep{Ginzburg_Syrovatskii1964} evaluated at $\Esyn$,
\beq
   \muabs\sim 10^2\,\frac{e}{B\sin\theta}\,\frac{1}{\gamma_e^4}\,\frac{dn_\pm}{d\gamma_e},
\eeq
where the numerical factor was evaluated using $d\ln n_\pm/d\ln\gamma_e\sim -2$.
The energy distribution of the cooling pairs is governed by the continuity equation in 
the energy space,
\beq
   \dot{\gamma}_e\,\frac{dn_\pm}{d\gamma_e}=\dot{n}, \qquad
   \dot{\gamma}_e=\frac{4}{3}\,\frac{\sT \Ueff}{m_ec}\,\gamma_e^2.
\eeq
Here $\dot{\gamma}_e$ is the cooling rate with 
$\Ueff=\sin^2\theta\; U_B+\zeta_{\rm KN}\Urad$, and $\dot{n}$ is the injection rate of 
pairs with Lorentz factors above $\gamma_e$. The injection rate may be written as 
\beq
   z_s \dot{n}=Y(\gamma_e)\,\frac{\fHE U_B\vrec}{m_ec^2},
\eeq
where $Y(\gamma_e)\sim 10^{-2}-10^{-1}$ is the pair yield above $\gamma_e$, and $z_s$ 
is the characteristic thickness of the layer where the secondary high-energy pairs are 
injected. 

The condition for self-absorption is $\muabs z_s/\sin\theta\sim 1$, where
$\theta$ is the injection angle of the pairs (and their synchrotron photons) with respect
to the magnetic field. This condition gives
\begin{eqnarray}
\nonumber
   \gabs &\sim& \left(\frac{10^2}{\sin^2\theta}\,\frac{e}{\sT B}\,Y\fHE\,\brec\,
         \frac{U_B}{\Ueff}\right)^{1/6} \\
           & \sim & \frac{15}{\sin^{1/3}\theta} 
    \left(\frac{\fHE}{0.1}\right)^{1/6}\left(\frac{M}{10M_\odot}\right)^{1/12},
\end{eqnarray}
where we substituted $B$ from \Eq~(\ref{eq:B}).
Self-absorption is particularly strong in magnetic flares in AGN, which 
have a typical black hole mass $M\sim 10^8M_\odot$.

Secondary particles with $\gamma_e<\gabs$ are cooled only by Compton scattering.
Besides blocking synchrotron cooling, self-absorption also provides means for 
thermalization of $e^\pm$ pairs \citep{Ghisellini+1998,Poutanen_Vurm2009}. 
In addition, at low energies, thermalization is assisted by Coulomb collisions. 
Detailed calculations of these processes show how the injection of high-energy particles 
leads to a Comptonized component extending to MeV energies \citep{Poutanen_Vurm2009}.
This component was not included in our simple Monte-Carlo simulation.
It should become visible in the spectrum of a magnetic flare at $E\simgt m_ec^2$
(where the emission from the chain bulk Comptonization is strongly suppressed).
It should, however, be cut off by $\gamma\gamma$ absorption at $E\gg 1$~MeV.
Any significant emission well above 1~MeV must 
originate at a larger distance from the black hole, most likely in a relativistic jet.

\subsection{Annihilation radiation}

The ratio of annihilation luminosity $\Lann$ to the total power $L$ dissipated (and 
radiated) by the reconnection layer is given by
\beq
  \frac{\Lann}{L}\approx \frac{2m_ec^2\dot{n}_{\rm ann} h}{U_B\vrec}
   \approx\frac{3\,\tauT^2}{16\,\brec^2\lB}\sim 10^{-2}.
\eeq
$\Lann$ is comparable to the luminosity $L_1$ in MeV photons that create $e^\pm$ 
pairs. In a flare with the typical $\lB\simgt 10^3$, the annihilation timescale is shorter 
than the residence time of particles in the reconnection layer (\Sect~2.3), so a large 
fraction of the created pairs annihilate, approaching the annihilation balance. 
Then the annihilation luminosity is related to the efficiency of pair creation and may 
be written as
\beq
  \frac{\Lann}{L}\approx Y\fHE.
\eeq

The annihilation photons have energies close to 511~keV in the rest frame of the
pair plasma. Their observed energies will be affected by gravitational and 
Doppler shifts. The reconnection layer has a size 
comparable to the Schwarzshild radius $r_g$, and the plasma in the layer has 
significant bulk speeds.
A large fraction of pairs annihilate inside plasmoids with fast random motions.
Note also that the entire flare region may have an interesting bulk speed controlled 
by the net flux of radiation away from the accretion disk \citep{Beloborodov1999c}.  
The pair plasma is light and its inertia is small, so it tends to assume an equilibrium 
speed along the magnetic field lines. This speed is such that the local net radiation 
flux, measured in the plasma rest frame, is perpendicular to the local magnetic field.


\section{Discussion}

\subsection{Radiative reconnection}

Reconnection in magnetic flares near luminous accreting black holes
occurs in the radiative regime, i.e. most of the dissipated energy promptly converts to radiation.
The plasma in the reconnection layer is still organized in the self-similar chain of plasmoids,
however radiative effects change the plasma state in three ways.
(1) Most of the plasma is cooled to a temperature comparable to the Compton temperature 
of the radiation field, $k\TC\sim 10$~keV.
(2) A large number of $e^\pm$ pairs are created in the reconnection layer and its 
scattering optical depth becomes comparable to unity.
(3) The bulk motions of pair-loaded plasmoids are limited by Compton drag. 
Monster plasmoids, which contain most of the plasma, are moving with mildly relativistic 
speeds. Small plasmoids move with high Lorentz factors $\gamma$, which 
are inversely proportional to the size (and optical depth) of the plasmoid, up to the 
maximum $\gamma=\sigma^{1/2}$.

A key feature of the radiative reconnection layer is that 
the plasma energy is dominated by the bulk motions of macroscopic plasmoids rather 
than thermal motions of individual particles. A fraction of particles reach high energies in 
intermittent acceleration events, in particular near the X-points, however they 
are quickly cooled and buried in the growing plasmoids of sizes $w\gg r_g/\lB$.
Mergers of cooled plasmoids are inefficient in pushing electrons to high energies.
Instead, dissipation in the reconnection chain mainly occurs through 
magnetic stresses stirring plasmoids against Compton drag. 
The distribution of the dissipated power $L$ over the plasmoid size $w$ is roughly 
estimated as $dL/dw\approx const$ at $w\ll r_g$. It peaks at large $w$, i.e. most of the 
power is deposited into large (mildly relativistic) plasmoids rather than small ones 
with high Lorentz factors.

Note that the ion component is not cooled, and ions should
form a broad energy distribution as observed in the PIC simulations without cooling. 
However, the $e^\pm$ loading of the flare reduces the energy budget of the ion 
component by the factor $Z^{-1}=\tauT^{ep}/\tauT\simlt\sigma^{-1}$.

A simplest flare model would assume that the magnetic energy advected into the 
reconnection layer is continually converted to heat. This is not happening --- the PIC 
simulations show no sign of continually heated Maxwellian plasma. Instead, 
nonthermal particles are accelerated at the X-points and deposited into young small 
plasmoids, which grow and cool down.
Effectively, a fraction $\fHE$ of the released magnetic energy is impulsively injected in 
the form of high-energy particles. The value of $\fHE$ may be measured  
in PIC simulations. Based on existing results  (L. Sironi, private communication), 
$\fHE\sim 0.1$ appears to be a reasonable estimate. The remaining fraction $1-\fHE$ 
is deposited into plasmoid bulk acceleration by magnetic stresses.

Thus, dissipation in magnetic flares occurs through two distinct modes: Compton 
drag on the chain and particle acceleration at X-points. This explains ``hybrid 
Comptonization'' observed in the hard-state of Cyg~X-1. It was previously 
modeled using a phenomenological picture of thermal+nonthermal dissipation 
\citep{Coppi1999} or pure nonthermal dissipation accompanied by electron thermalization 
through synchrotron self-absorption \citep{Poutanen_Vurm2009,Poutanen_Veledina2014}. 
The latter model fits the data only with a relatively low magnetic field (and a large size 
of the source), which is energetically inconsistent with magnetic flares. 
This difficulty is resolved when plasma is not required to sustain $kT_e\approx 100$~keV, 
and instead the observed hard X-rays are generated by the chain Comptonization.

\medskip

\subsection{X-ray spectrum}

Figure~2 suggests that radiative reconnection is a natural producer of hard X-ray
spectra observed in X-ray binaries and AGN, with no need to assume a hot 
Maxwellian plasma. In agreement with observations, the chain Comptonization 
produces a hard X-ray spectrum with a sharp cutoff around 100~keV. 
The model calculated in \Sect~4.1 has the same number of parameters as thermal 
Comptonization ($T_s$, $A$, $\tauT$), and can be used to directly fit the observed spectra.

The spectral cutoff is located below $m_ec^2=511$~keV for 
the reasons discussed in \Sect~4.1. Its exact position depends on the optical 
depth of the reconnection layer $\tauT$ (Figure~2), which is regulated by pair creation.
The expected range of the flare compactness $3\times 10^2\simlt\lB\simlt 10^4$ 
approximately corresponds to $0.5\simlt \tauT\simlt 3$, which results in variations of 
the cutoff position between 200 and 40~keV. A similar range is observed in the hard states
of accreting black holes in X-ray binaries, e.g. Cyg~X-1 and GX~339-4, and AGN.
 
The Comptonized hard X-ray spectrum forms because the plasmoid chain is exposed to 
soft radiation with luminosity $L_s$, which can be much smaller than the flare power $L$. 
The spectral slope of chain emission is controlled by the Compton amplification factor 
$A=L/L_s$. $L_s$ can be generated by the flare itself in two ways: 
(1) part of the flare radiation is intercepted by cold gas in the accretion disk and 
reprocessed into soft photons, and 
(2) soft synchrotron radiation is produced by the electrons accelerated in the reconnection 
layer. 
The observed spectral slope of Cyg~X-1 (photon index $\Gamma\approx 1.6$)
requires $A\sim 10$, consistent with a low $L_s\sim 0.1 L$. 

The efficiency of reprocessing depends on the geometry of the disk+corona configuration, 
the albedo of the disk surface ionized by the flare, and the anisotropy of the flare emission. 
The flaring $e^\pm$ plasma is preferentially ejected away from the disk, which makes the 
emission strongly anisotropic and reduces its reprocessing/reflection 
\citep{Beloborodov1999c,Malzac+2001}.
Uncertainties in the corona configuration may eventually be resolved by observations.
In particular, the iron $K_\alpha$ line in the reflected spectrum 
provides a useful tool to study the innermost region of the accretion disk
\citep{Fabian2016}. Flares must be accompanied by X-ray reverberation due to reflection,
which has been observed in AGN on timescales comparable to $r_g/c$
\citep{Kara+2016}. Future observations with rich photon statistics on short timescales 
may clarify the role of reprocessing in photon supply to the reconnection regions.

In addition to chain Comptonization, magnetic flares generate high-energy particles, 
which receive a fraction $\fHE$ of the released power.
Their energy partially converts to soft synchrotron radiation and partially feeds 
an IC cascade, which must form a spectral tail extending to $\simgt 1$~MeV. 
The high-energy component needs more detailed calculations similar to 
those in \citet{Poutanen_Vurm2009}. It may explain the observed spectral tail 
sticking out at $E\simgt m_ec^2$ in Cyg~X-1 \citep{McConnell+2002}.  
The composite spectrum is a natural result of the 
hybrid Comptonization by the chain and accelerated particles, however, 
the spectrum at high energies may be further complicated by additional 
contributions from the jet \citep{Zdziarski+2016}.

Sustaining the optical depth $\tauT\sim 1$ through pair creation implies a significant 
rate of $e^\pm$ annihilation, which should produce a spectral feature around 511~keV.
Its luminosity is comparable to 1\% of the flare power (\Sect~4.3).
The annihilation line may be hard to detect because it is shifted and broadened by 
the gravitational and Doppler effects. 
There is, however, some evidence for pair plasma near accreting black holes.
A broad, variable annihilation feature was seen in the recent outburst of V404~Cyg
\citep{Siegert+2016} and previously reported in a few other black-hole candidates 
\citep{Goldwurm+1992,Sunyaev+1992}.
No annihilation feature has yet been identified in Cyg~X-1.
The detailed shape of the spectrum around 0.5~MeV is difficult to measure, 
because of the relatively low flux and a modest detector sensitivity in this energy band.

Magnetic flares with $10^3<\lB<10^4$ are expected to occur in the powerful,  magnetically dominated corona of the accretion disk or at the jet base. In addition, the described picture of radiative reconnection should apply to a broader class of magnetic flares. Flares with $10\ll\lB\ll 10^3$ are still radiative, however they create less $e^\pm$ plasma, so the region stays optically thin. Then the emission is expected to have a more extended nonthermal spectrum. This may occur in a weak corona associated with a soft spectral state, generating an extended tail in the radiation spectrum. Low-$\lB$ flares can also happen in relativistic jets at some distance from the black hole.

\acknowledgements
I am grateful to L. Sironi and C. Lundman for discussions and comments on the 
manuscript. This work was supported by a grant from the Simons Foundation 
(\#446228, Andrei Beloborodov). 



\bibliography{ms.bbl}

\begin{thebibliography}{}
\expandafter\ifx\csname natexlab\endcsname\relax\def\natexlab#1{#1}\fi

\bibitem[{{Beloborodov}(1999{\natexlab{a}})}]{Beloborodov1999a}
{Beloborodov}, A.~M. 1999{\natexlab{a}}, in Astronomical Society of the Pacific
  Conference Series, Vol. 161, High Energy Processes in Accreting Black Holes,
  ed. J.~{Poutanen} \& R.~{Svensson}, 295

\bibitem[{{Beloborodov}(1999{\natexlab{b}})}]{Beloborodov1999b}
{Beloborodov}, A.~M. 1999{\natexlab{b}}, \mnras, 305, 181

\bibitem[{{Beloborodov}(1999{\natexlab{c}})}]{Beloborodov1999c}
---. 1999{\natexlab{c}}, \apjl, 510, L123

\bibitem[{{Cerutti} {et~al.}(2014){Cerutti}, {Werner}, {Uzdensky}, \&
  {Begelman}}]{Cerutti+2014}
{Cerutti}, B., {Werner}, G.~R., {Uzdensky}, D.~A., \& {Begelman}, M.~C. 2014,
  \apj, 782, 104

\bibitem[{{Coppi}(1999)}]{Coppi1999}
{Coppi}, P.~S. 1999, in Astronomical Society of the Pacific Conference Series,
  Vol. 161, High Energy Processes in Accreting Black Holes, ed. J.~{Poutanen}
  \& R.~{Svensson}, 375

\bibitem[{{Fabian}(2016)}]{Fabian2016}
{Fabian}, A.~C. 2016, Astronomische Nachrichten, 337, 375

\bibitem[{{Fabian} {et~al.}(2015){Fabian}, {Lohfink}, {Kara}, {Parker},
  {Vasudevan}, \& {Reynolds}}]{Fabian+2015}
{Fabian}, A.~C., {Lohfink}, A., {Kara}, E., {et~al.} 2015, \mnras, 451, 4375

\bibitem[{{Galeev} {et~al.}(1979){Galeev}, {Rosner}, \& {Vaiana}}]{Galeev+1979}
{Galeev}, A.~A., {Rosner}, R., \& {Vaiana}, G.~S. 1979, \apj, 229, 318

\bibitem[{{Ghisellini} {et~al.}(1998){Ghisellini}, {Haardt}, \&
  {Svensson}}]{Ghisellini+1998}
{Ghisellini}, G., {Haardt}, F., \& {Svensson}, R. 1998, \mnras, 297, 348

\bibitem[{{Gierli{\'n}ski} \& {Zdziarski}(2003)}]{Gierlinski_Zdziarski2003}
{Gierli{\'n}ski}, M., \& {Zdziarski}, A.~A. 2003, \mnras, 343, L84

\bibitem[{{Ginzburg} \& {Syrovatskii}(1964)}]{Ginzburg_Syrovatskii1964}
{Ginzburg}, V.~L., \& {Syrovatskii}, S.~I. 1964, {The Origin of Cosmic Rays}

\bibitem[{{Goldwurm} {et~al.}(1992){Goldwurm}, {Ballet}, {Cordier}, {Paul},
  {Bouchet}, {Roques}, {Barret}, {Mandrou}, {Sunyaev}, {Churazov}, {Gilfanov},
  {Dyachkov}, {Khavenson}, {Kovtunenko}, {Kremnev}, \&
  {Sukhanov}}]{Goldwurm+1992}
{Goldwurm}, A., {Ballet}, J., {Cordier}, B., {et~al.} 1992, \apjl, 389, L79

\bibitem[{{Guilbert} {et~al.}(1983){Guilbert}, {Fabian}, \&
  {Rees}}]{Guilbert+1983}
{Guilbert}, P.~W., {Fabian}, A.~C., \& {Rees}, M.~J. 1983, \mnras, 205, 593

\bibitem[{{Guo} {et~al.}(2016){Guo}, {Li}, {Li}, {Daughton}, {Zhang},
  {Lloyd-Ronning}, {Liu}, {Zhang}, \& {Deng}}]{Guo+2016}
{Guo}, F., {Li}, X., {Li}, H., {et~al.} 2016, \apjl, 818, L9

\bibitem[{{Huang} \& {Bhattacharjee}(2012)}]{Huang_Bhattacharjee2012}
{Huang}, Y.-M., \& {Bhattacharjee}, A. 2012, Physical Review Letters, 109,
  265002

\bibitem[{{Kagan} {et~al.}(2016){Kagan}, {Nakar}, \& {Piran}}]{Kagan+2016}
{Kagan}, D., {Nakar}, E., \& {Piran}, T. 2016, \apj, 826, 221

\bibitem[{{Kara} {et~al.}(2016){Kara}, {Alston}, {Fabian}, {Cackett}, {Uttley},
  {Reynolds}, \& {Zoghbi}}]{Kara+2016}
{Kara}, E., {Alston}, W.~N., {Fabian}, A.~C., {et~al.} 2016, \mnras, 462, 511

\bibitem[{{Loureiro} {et~al.}(2012){Loureiro}, {Samtaney}, {Schekochihin}, \&
  {Uzdensky}}]{Loureiro+2012}
{Loureiro}, N.~F., {Samtaney}, R., {Schekochihin}, A.~A., \& {Uzdensky}, D.~A.
  2012, Physics of Plasmas, 19, 042303

\bibitem[{{Lyubarsky}(2005)}]{Lyubarsky2005}
{Lyubarsky}, Y.~E. 2005, \mnras, 358, 113

\bibitem[{{Malzac} {et~al.}(2001){Malzac}, {Beloborodov}, \&
  {Poutanen}}]{Malzac+2001}
{Malzac}, J., {Beloborodov}, A.~M., \& {Poutanen}, J. 2001, \mnras, 326, 417

\bibitem[{{McConnell} {et~al.}(2002){McConnell}, {Zdziarski}, {Bennett},
  {Bloemen}, {Collmar}, {Hermsen}, {Kuiper}, {Paciesas}, {Phlips}, {Poutanen},
  {Ryan}, {Sch{\"o}nfelder}, {Steinle}, \& {Strong}}]{McConnell+2002}
{McConnell}, M.~L., {Zdziarski}, A.~A., {Bennett}, K., {et~al.} 2002, \apj,
  572, 984

\bibitem[{{Melzani} {et~al.}(2014){Melzani}, {Walder}, {Folini},
  {Winisdoerffer}, \& {Favre}}]{Melzani+2014}
{Melzani}, M., {Walder}, R., {Folini}, D., {Winisdoerffer}, C., \& {Favre},
  J.~M. 2014, \aap, 570, A112

\bibitem[{{Odell}(1981)}]{Odell1981}
{Odell}, S.~L. 1981, \apjl, 243, L147

\bibitem[{{Parfrey} {et~al.}(2013){Parfrey}, {Beloborodov}, \&
  {Hui}}]{Parfrey+2013}
{Parfrey}, K., {Beloborodov}, A.~M., \& {Hui}, L. 2013, \apj, 774, 92

\bibitem[{{Parfrey} {et~al.}(2015){Parfrey}, {Giannios}, \&
  {Beloborodov}}]{Parfrey+2015}
{Parfrey}, K., {Giannios}, D., \& {Beloborodov}, A.~M. 2015, \mnras, 446, L61

\bibitem[{{Petropoulou} {et~al.}(2016){Petropoulou}, {Giannios}, \&
  {Sironi}}]{Petropoulou+2016}
{Petropoulou}, M., {Giannios}, D., \& {Sironi}, L. 2016, \mnras, 462, 3325

\bibitem[{{Phinney}(1982)}]{Phinney1982}
{Phinney}, E.~S. 1982, \mnras, 198, 1109

\bibitem[{{Poutanen} \& {Veledina}(2014)}]{Poutanen_Veledina2014}
{Poutanen}, J., \& {Veledina}, A. 2014, \ssr, 183, 61

\bibitem[{{Poutanen} \& {Vurm}(2009)}]{Poutanen_Vurm2009}
{Poutanen}, J., \& {Vurm}, I. 2009, \apjl, 690, L97

\bibitem[{{Romanova} {et~al.}(1998){Romanova}, {Ustyugova}, {Koldoba},
  {Chechetkin}, \& {Lovelace}}]{Romanova+1998}
{Romanova}, M.~M., {Ustyugova}, G.~V., {Koldoba}, A.~V., {Chechetkin}, V.~M.,
  \& {Lovelace}, R.~V.~E. 1998, \apj, 500, 703

\bibitem[{{Rybicki} \& {Lightman}(1979)}]{Rybicki_Lightman1979}
{Rybicki}, G.~B., \& {Lightman}, A.~P. 1979, {Radiative processes in
  astrophysics}

\bibitem[{{Siegert} {et~al.}(2016){Siegert}, {Diehl}, {Greiner}, {Krause},
  {Beloborodov}, {Bel}, {Guglielmetti}, {Rodriguez}, {Strong}, \&
  {Zhang}}]{Siegert+2016}
{Siegert}, T., {Diehl}, R., {Greiner}, J., {et~al.} 2016, \nat, 531, 341

\bibitem[{{Sikora} {et~al.}(1996){Sikora}, {Sol}, {Begelman}, \&
  {Madejski}}]{Sikora+1996}
{Sikora}, M., {Sol}, H., {Begelman}, M.~C., \& {Madejski}, G.~M. 1996, \mnras,
  280, 781

\bibitem[{{Sironi} {et~al.}(2016){Sironi}, {Giannios}, \&
  {Petropoulou}}]{Sironi+2016}
{Sironi}, L., {Giannios}, D., \& {Petropoulou}, M. 2016, \mnras, 462, 48

\bibitem[{{Sironi} \& {Spitkovsky}(2014)}]{Sironi_Spitkovsky2014}
{Sironi}, L., \& {Spitkovsky}, A. 2014, \apjl, 783, L21

\bibitem[{{Stern} {et~al.}(1995){Stern}, {Poutanen}, {Svensson}, {Sikora}, \&
  {Begelman}}]{Stern+1995}
{Stern}, B.~E., {Poutanen}, J., {Svensson}, R., {Sikora}, M., \& {Begelman},
  M.~C. 1995, \apjl, 449, L13

\bibitem[{{Sunyaev} {et~al.}(1992){Sunyaev}, {Churazov}, {Gilfanov},
  {Dyachkov}, {Khavenson}, {Grebenev}, {Kremnev}, {Sukhanov}, {Goldwurm},
  {Ballet}, {Cordier}, {Paul}, {Denis}, {Vedrenne}, {Niel}, \&
  {Jourdain}}]{Sunyaev+1992}
{Sunyaev}, R., {Churazov}, E., {Gilfanov}, M., {et~al.} 1992, \apjl, 389, L75

\bibitem[{{Svensson}(1984)}]{Svensson1984}
{Svensson}, R. 1984, \mnras, 209, 175

\bibitem[{{Svensson}(1987)}]{Svensson1987}
---. 1987, \mnras, 227, 403

\bibitem[{{Tchekhovskoy} {et~al.}(2011){Tchekhovskoy}, {Narayan}, \&
  {McKinney}}]{Tchekhovskoy+2011}
{Tchekhovskoy}, A., {Narayan}, R., \& {McKinney}, J.~C. 2011, \mnras, 418, L79

\bibitem[{{Uzdensky} {et~al.}(2010){Uzdensky}, {Loureiro}, \&
  {Schekochihin}}]{Uzdensky+2010}
{Uzdensky}, D.~A., {Loureiro}, N.~F., \& {Schekochihin}, A.~A. 2010, Physical
  Review Letters, 105, 235002

\bibitem[{{Werner} {et~al.}(2016){Werner}, {Uzdensky}, {Cerutti}, {Nalewajko},
  \& {Begelman}}]{Werner+2016}
{Werner}, G.~R., {Uzdensky}, D.~A., {Cerutti}, B., {Nalewajko}, K., \&
  {Begelman}, M.~C. 2016, \apjl, 816, L8

\bibitem[{{Zdziarski} \& {Gierli{\'n}ski}(2004)}]{Zdziarski_Gierlinski2004}
{Zdziarski}, A.~A., \& {Gierli{\'n}ski}, M. 2004, Progress of Theoretical
  Physics Supplement, 155, 99

\bibitem[{{Zdziarski} {et~al.}(2016){Zdziarski}, {Malyshev}, {Chernyakova}, \&
  {Pooley}}]{Zdziarski+2016}
{Zdziarski}, A.~A., {Malyshev}, D., {Chernyakova}, M., \& {Pooley}, G.~G. 2016,
  ArXiv e-prints, arXiv:1607.05059

\end{thebibliography}

\end{document}